\documentclass[a4paper]{jpconf}
\usepackage{amsfonts}
\usepackage{amssymb}

\begin{document}
\title{Genuine converging solution of self-consistent field equations for extended many-electron systems}
\author{A.Ya. Shul'man}
\address{Institute of Radio Engineering and Electronics of the RAS, 125009 Moscow, Russia}
\ead{ash@cplire.ru}
\begin{abstract}
Calculations of the ground state of inhomogeneous many-electron systems
involve a solving of the Poisson equation for Coulomb potential and the
Schr\"{o}dinger equation for single-particle orbitals. Due to nonlinearity and
complexity this set of equations, one believes in the iterative method for the
solution that should consist in consecutive improvement of the potential and
the electron density until the self-consistency is attained. Though this
approach exists for a long time there are two grave problems accompanying its
implementation to infinitely extended systems.

The first of them is related with the Poisson equation and lies in possible
incompatibility of the boundary conditions for the potential with the electron
density distribution renewed by means of the Schr\"{o}dinger equation.
Rigorously speaking, it means the iteration process cannot be continued. The
resolution of this difficulty is proposed for both infinite conducting systems
in jellium approximation and periodic structures. It provides the existence of
self-consistent solution for the potential at every iteration step due to
realization of a screening effect.

The second problem results from the existence of continuous spectrum of
Hamiltonian eigenvalues for unbounded systems. It needs to have a definition
of Hilbert space basis with eigenfunctions of continuous spectrum as elements,
which would be convenient in numerical applications. It is proposed to insert
a limiting transition into the definition of scalar product specifying the
Hilbert space. It provides self-adjointness of Hamiltonian and, respectively,
the orthogonality of eigenfunctions corresponding to the different
eigenvalues. In addition, it allows to normalize them effectively to
delta-function and to prove the orthogonality of the 'right' and 'left'
eigenfunctions belonging to twofold degenerate eigenvalues.
\end{abstract}
\section{Problem definition}
\label{Introd}
Ground-state calculations of inhomogeneous
many-electron systems involve generally a solving of the Poisson
equation for averaged Coulomb potential $u(\mathbf{r})$ at given
spatial electron density $n(\mathbf{r})$ and the Schr\"{o}dinger
equation for single-particle orbitals $\psi
_{\mathrm{E}}(\mathbf{r})$\ in the potential $u_{eff}$ accounting by
some approximation for the difference between $u(\mathbf{r})$ and
the microscopic local field. In the density functional theory the
corresponding set of Kohn-Sham equations for the spin-unpolarized
electron gas has the form (in atomic units $|e|=m=\hbar=1$)
\begin{equation}
-\frac{1}{2}\nabla^{2}\Psi_{E}(\mathbf{r})+u_{\mathrm{eff}}(\mathbf{r}%
)\Psi_{E}(\mathbf{r})=E\Psi_{E}(\mathbf{r})\label{Schroed}%
\end{equation}%
\begin{equation}
u_{\mathrm{eff}}(\mathbf{r})=u(\mathbf{r})+u_{\mathrm{xc}}(\mathbf{r}%
)\label{Eff-pot}%
\end{equation}%
\begin{equation}
\nabla^{2}u(\mathbf{r})=4\pi(N_{\mathrm{+}}(\mathbf{r})-n(\mathbf{r}%
))\label{Poisson-gen}%
\end{equation}%
\begin{equation}
n(\mathbf{r})=2\sum_{E\leqslant E_{F}}|\Psi_{E}|^{2}(\mathbf{r}%
)\label{n(r)-gen}%
\end{equation}
Here $E$ is the energy eigenvalue of the single-particle
Hamiltonian, $E_{F}$ is the Fermi energy of the electrons,
$N_{\mathrm{+}}$ is the density of the positive background, $u$ is
the Coulomb potential energy of the electron, and $u_{\mathrm{xc}}$
is the exchange-correlation potential energy, which is assumed in
the local-density approximation $u_{\mathrm{xc}}(\mathbf{r})\equiv
u_{\mathrm{xc}}\left[  n(\mathbf{r})\right]  $.

Due to nonlinearity and complexity of this set of equations, one believes in
the iterative solution procedure that should consist in consecutive
improvement of $u(\mathbf{r})$ and $n(\mathbf{r})$ until the self-consistency
is attained. While this approach exists for a long time and as if were used in
many articles there are two grave problems accompanying its applications and
pertinent to very background of the iterative method in the case of infinitely
extended many-electron systems.

The first of them is related with Poisson equation and lies in possible
incompatibility of the boundary conditions for $u(\mathbf{r})$ with the
distribution $n(\mathbf{r})$ renewed by means of the Schr\"{o}dinger equation
solutions and substituted to the right-hand side of the Poisson equation
(\ref{Poisson-gen}). Rigorously speaking, it means the iteration process
cannot be continued. To manage this difficulty there were several empirical
techniques suggested (see e.c. \cite{Liebsch97}) but their shortcomings are
either lack of iteration convergence and a transfer to some kind of variation
solution instead (see \cite{L-K70PRB4555}, Appendix B), or an appearance of
solution instability \cite{Ferr-Sm85PRB3427} or a change in the positive
charge distribution \cite{Liebsch97} that violates the derivation conditions
of self-consistent field equations as variational equations . In the case of
extended but finite systems, the effect may result in the non-physical growth
of the electric filed far away from the inhomogeneity region
\cite{Frens90RMP745}. The compliance of obtained solution with true one is an
open question in all cases. The suggested approach to deal with this problem
are described in Section \ref{Poiss-iter sec} for systems with Fermi level
given at the infinity and in Section \ref{finite number sec} for systems with
given number of electrons.

The second problem results from the existence of continuous spectrum of
Hamiltonian eigenvalues for unbounded systems. It is necessary to have such a
definition of Hilbert space with eigenfunctions of continuous spectrum as
elements, which would be convenient in numerical applications. A limiting
transition to Hamiltonian operator with continuous spectrum by means of the
unlimited increase in system size is used to build up mathematically the
Hilbert space of singular self-adjoined operators (see e.c. \cite{Lev-S70}).
However, this way is practically inappropriate for inhomogeneous systems.

It is suggested here to introduce a limiting transition into the definition of
the scalar product specifying the Hilbert space. The adopted form of the
scalar product provides self-adjointness of Hamiltonian operator and,
respectively, the orthogonality of all eigenfunctions corresponding to the
different eigenvalues. In addition, it allows to normalize them effectively to
delta-function and to prove the orthogonality of the ''right'' and ''left''
current-carried eigenfunctions belonging to twofold degenerate eigenvalues.
This is particularly of the essence for the problem of tunneling through a
self-consistent barrier considered as an example in Section \ref{Schred-Hilb
Sec}. Also this approach is implemented to the Bloch wave functions of
periodic solids in Section \ref{finite number sec}.

\section{Poisson equation and iteration algorithm}
\label{Poiss-iter sec}
Let us illustrate the neutrality problem with
one-dimensional Poisson equation given on semi-axis $z\in\lbrack0,\infty)$:%
\begin{equation}
u^{\prime\prime}(z)=4\pi\rho(z),\,\,\rho=N_{\mathrm{+}}%
(z)-n(z)\label{Poiss-1D}%
\end{equation}
The simple integration results in%
\begin{equation}
u^{\prime}(z)=u^{\prime}(0)+4\pi\int_{0}^{z}dz_{1}\rho(z_{1}%
),\label{1D u' solution}%
\end{equation}%
\begin{equation}
u(z)=u(0)+zu^{\prime}(z)-4\pi\int_{0}^{z}dz_{1}z_{1}\rho(z_{1}%
).\label{1D u solution}%
\end{equation}
The finiteness condition of $u(\infty)<\infty$ requires
$\lim_{z\rightarrow \infty}zu^{\prime}(z)=0$ and
\begin{equation}
4\pi\int_{0}^{\infty}dz_{1}\rho(z_{1})=-u^{\prime}(0),\label{charge cond}%
\end{equation}%
\begin{equation}
4\pi\int_{0}^{\infty}dz_{1}z_{1}\rho(z_{1})=u(0)-u(\infty)\label{dipole cond}%
\end{equation}
Assuming an energy scale such that $u(\infty)=0$ it is easily to see
the existence of the direct relationship between the total charge
and a boundary condition for the electric field $u^{\prime}(0)$ in
Eq.(\ref{1D u' solution}) or between the total dipole moment and a
boundary condition for the potential $u(0)$ in Eq.(\ref{1D u
solution}). However, the electron density $n^{(i)}(\mathbf{r})$ in
Eq.(\ref{n(r)-gen}) obtained after the solution of the
Schr\"{o}dinger Eq.(\ref{Schroed}) at an $i$-th iteration step may
not satisfy the imposed boundary conditions for the Poisson equation
as usually it takes place. In this case there is no possibility to
solve the Eq.(\ref{Poiss-1D}) and the iteration procedure should be
stopped.

In the presented approach this difficulty is removed by partition of full
density $n(\mathbf{r})$ in two terms%
\begin{equation}
n(z)=n_{\mathrm{ind}}\left[  u(z)\right]  +n_{\mathrm{Q}}(z), \label{N=Ni+Nq}%
\end{equation}
where $n_{\mathrm{ind}}$ is defined as a function of the Coulomb potential $u$
via its relation to the effective potential $u_{\mathrm{eff}}%
=u(z)+u_{\mathrm{xc}}\left[  n(z)\right]  $ by the known quasi-classical
expression%
\begin{equation}
n_{\mathrm{ind}}\left(  u,n\right)  =\frac{2^{3/2}}{3\pi^{2}}\left[
E_{F}-u_{\mathrm{eff}}(z,n(z))\right]  ^{3/2} \label{Ni-quasicl}%
\end{equation}
and%
\begin{equation}
n_{\mathrm{Q}}(z)=n(z)-n_{\mathrm{ind}}(z) \label{Nq-definit}%
\end{equation}
is named as the quantum correction.\ Using the definition (\ref{N=Ni+Nq}), the
Poisson equation (\ref{Poiss-1D}) can be rewritten in the form%
\begin{equation}
u^{\prime\prime}+4\pi n_{\mathrm{ind}}[u,n_{\mathrm{Q}}]=4\pi(N_{\mathrm{+}%
}-n_{\mathrm{Q}}(z)). \label{Poiss+Ni}%
\end{equation}
If the pair functions $n(z)$ and $u(z)$ is the true self-consistent
solution of the problem (\ref{Schroed}-\ref{n(r)-gen}) then the
Eq.(\ref{Poiss+Ni}) is the simply rearranged Eq.(\ref{Poiss-1D}).
However, the Eq.(\ref{Poiss+Ni}) is much more appropriate for the
iterative procedure since, due to the screening effect, the induced
electron density $n_{\mathrm{ind}}^{(i)}$ depending on the unknown
Coulomb potential provides the existence of self-consistent solution
$u^{(i)}(\mathbf{r})$ at every iteration step and any possible
spatial dependence of right-hand side of Eq.(\ref{Poiss+Ni}). It is
useful to note that the expression (\ref{Ni-quasicl}) is a good
approximation of the solution of Schr\"{o}dinger equation for the
smooth part of the $u_{\mathrm{eff}}$ and produces the true
screening of the long-range part of the Coulomb potential since the
$n_{\mathrm{ind}}$.is found simultaneously with $u$ in the course of
self-consistent solution of the Poisson equation. The rest
short-range variations of the density $n(z)$ are exactly described
by n$_{\mathrm{Q}}(z)$, which is to be found in the usual iterative
cycle after the solution of Schr\"{o}dinger equation. This is the
root cause of the algorithm efficiency. It will be convenient to
designate the Eq. (\ref{Poiss+Ni}) as self-consistent Poisson
equation.

The full iteration algorithm in the case of the one-dimensional inhomogeneity
of the charge distribution can be described now as following%
\begin{equation}
i=0,1,...;\,\,\,n_{\mathrm{Q}}^{(0)}=0,
\end{equation}%
\begin{equation}
u^{\prime\prime(i)}+4\pi n_{\mathrm{ind}}\left[  u^{(i)},n_{\mathrm{Q}}%
^{(i)}\right]  =4\pi(N_{\mathrm{+}}(z)-n_{\mathrm{Q}}^{(i)}%
(z)),\label{Poiss-it}%
\end{equation}%
\begin{equation}
n_{\mathrm{s}}^{(i)}(z)=n_{\mathrm{ind}}\left[  u^{(i)},n_{\mathrm{Q}}%
^{(i)}\right]  +n_{\mathrm{Q}}^{(i)}(z);\,\,u_{\mathrm{eff}}^{(i)}%
(z)=u^{(i)}(z)+u_{\mathrm{xc}}\left(  n_{\mathrm{s}}^{(i)}(z)\right)
,\label{Ns-it}%
\end{equation}%
\begin{equation}
\frac{1}{2}\psi_{k}^{\prime\prime(i)}(z)+\left(  \frac{1}{2}k^{2}%
+u_{\mathrm{eff}}(\infty)-u_{\mathrm{eff}}^{(i)}(z)\right)  \psi_{k}%
^{(i)}(z)=0,\label{Schred-it}%
\end{equation}%
\begin{equation}
E=\frac{1}{2}(k^{2}+\mathbf{k}_{||}^{2}+u_{\mathrm{eff}}(\infty));\,\,\Psi
_{E}^{(i)}(\mathbf{r})=\frac{1}{2\pi}\exp(i\mathbf{k}_{||}\mathbf{r}_{||}%
)\psi_{k}^{(i)}(z);\,\,n^{(i)}(z)=2\sum_{E\leqslant E_{F}}|\Psi_{E}^{(i)}%
|^{2}(\mathbf{r}),\label{N-it}%
\end{equation}%
\begin{equation}
n_{\mathrm{Q}}^{(i+1)}(z)=n^{(i)}(z)-n_{\mathrm{ind}}\left[  u^{(i)}%
(z),n^{(i)}(z)\right]  .\label{Nq-next}%
\end{equation}
Here $n_{\mathrm{s}}^{(i)}(z)$ is the electron density that is
self-consistent with the Coulomb potential $u^{(i)}(z)$ at the given
quantum correction density n$_{\mathrm{Q}}^{(i)}(z))$. The
self-consistent Poisson equation (\ref{Poiss-it}) is solved as the
boundary problem and the Schr\"{o}dinger equation (\ref{Schred-it})
is solved as the Cauchy problem.

It needs to say a few words relative to the speciality of the
quasi-classical expression (\ref{Ni-quasicl}) for $n_{\mathrm{ind}}$
in the case of density functional approach. Because
$u_{\mathrm{xc}}$ depends itself on the electron density the
Eq.(\ref{Ni-quasicl}) determines $n_{\mathrm{ind}}$ as the implicit
function of the Coulomb potential $u$ when $n_{\mathrm{Q}}$ is
known. This function has the physical meaning only under condition
$\partial n_{\mathrm{ind}}/\partial
u<0$, which is the stability condition for solutions of the Eq.(\ref{Poiss-it}%
) (see a discussion in \cite{ASH2001}).

The validity of the described method for semi-infinite electron systems was
examined by calculations of surface properties of simple metals and quantum
corrections to the capacity of barrier structures \cite{Sh-P02}. The expected
convergency of the iteration procedure has been obtained and the true
self-consistency of the solution has been successfully checked by the
Budd-Vannimenus criterion \cite{Budd-Van73}.

\section{Schr\"{o}dinger equation and Hilbert space}
\label{Schred-Hilb Sec}Difficulties of calculations of wave
functions of the continuous spectrum are also present in problems of
type of surface properties of metals where one has deal with the
semi-infinite inhomogeneous electron gas. Attempts to replace the
unbounded system by a system of finite size give rise to serious
complications both in analytical and in numerical computations (see
e.c. \cite{Paasch-Hiet83}). However, it is more instructive to
analyze here the problem by an example of the tunneling in
many-electron system.

In this case the questions related to the eigenfunctions pertinent to the
continuous spectrum of the Schr\"{o}dinger equation (\ref{Schroed}) can be
considered with a fair degree of details and usefulness. In such a system the
self-consistent Coulomb and exchange-correlation potentials effect inevitably
on the shape of the tunnel barrier. The resultant contribution to tunnel
current-voltage characteristics can vary from insignificant, as in
metal-insulator-metal junctions, up to decisive, as in the Schottky-barrier
metal-semiconductor junctions (see \cite{ASH2001} and references therein). The
essential dependence of the transparency of self-consistent barrier on the
energy of tunneling electrons and a reconstruction of the barrier with the
applied bias voltage make in principal impossible the use of the tunnel
Hamiltonian approach to describe such systems. Therefore it is necessary to
formulate some regular scheme of tunnel current calculations that would form
also a basis for the numerical realization of self-consistent solution.

\subsection{Scalar product and orthonormal basis in single-particle continuous spectrum}
\label{basis}
Let two parts of system (left - \textrm{L}, right -
\textrm{R}) occupy half-space $z<0$ (metal) and $z>0$ (semiconductor
with a degenerate electron gas and the Schottky barrier),
respectively. The effective potential energy $V(z)$ of electrons is
considered as independent of coordinates in the
$(x,y)$ interface plane. Then the single-particle Hamilton operator is:%
\begin{equation}
\hat{H}=\hat{T}+V(z), \label{Hamilton}%
\end{equation}
where $\hat{T}$ is the kinetic energy operator. Let us assume following
conditions for the asymptotes of $V(z)$%
\begin{equation}
V(z)\rightarrow V^{L},\,z\rightarrow-\infty;\,\,\,V(z)\rightarrow
0,\,z\rightarrow\infty. \label{V(z) def}%
\end{equation}
In the case of metal-semiconductor junctions one can accept $V(z)\equiv
V^{L}<0$ for $z\leq0$. Due to constraints (\ref{V(z) def}) the eigenfunctions
of continuous spectrum $\hat{H}$ have an oscillating behavior at the left or
both infinities depending on the relation between the $V^{L}$ and the energy
eigenvalue $E$. The second type eigenstates only give the contribution to the
current. Let $k\geq0$ is the wavevector of wave function oscillations at
$z\rightarrow\infty$ and $q\geq0$ is the same at $z\rightarrow-\infty$. The
eigenfunctions obey the equation%
\begin{equation}
\hat{H}\Psi_{E}(x,y,z)=E\Psi_{E}(x,y,z) \label{H-E}%
\end{equation}
and, in view of the translation invariance along the interface, they can be
taken in the form%
\begin{equation}
\Psi_{E}(x,y,z)=C_{k}\psi_{k}(z)\exp(i\mathbf{k}_{||}\mathbf{r}_{||})/2\pi,
\label{Psi-gen}%
\end{equation}
where $E^{R}=(k^{2}+\mathbf{k}_{||}^{2})/2$ is the energy spectrum of
electrons in the bulk of semiconductor.

The wave functions of continuous spectrum should be normalized to
the $\delta $-function of quantum numbers. The Eq.(\ref{Psi-gen})
provides already the normalization to
$\delta(\mathbf{k}_{||}-\mathbf{k}_{||}^{\prime})$ in the lateral
plane. This result ($2\pi$ in the denominator) is usually obtained
by means of Born-Karman periodic boundary condition in the
normalization box. Evidently, the system under consideration has no
periodicity in $z$ direction. To determine the constant $C$ and to
avoid cumbersome calculations of eigenfunctions for a finite-size
system mentioned in Section \ref{Introd} let
us define the scalar product of the eigenfunctions by%
\begin{equation}
\left\langle \psi_{k}|\psi_{k_{1}}\right\rangle =\lim_{\epsilon\rightarrow
0}\int_{-\infty}^{\infty}dz\exp(-\epsilon|z|)\psi_{k}^{\ast}(z)\psi_{k_{1}%
}(z)\label{scalar-prod}%
\end{equation}
with the natural definition of the eigenfunction norm by
\begin{equation}
\left\|  \psi_{k}\right\|  ^{2}=\lim_{k_{1}\rightarrow k}\left\langle \psi
_{k}|\psi_{k_{1}}\right\rangle .\label{norm-def}%
\end{equation}

It is easily to check that the operator $\hat{T}$ \ in Eq. (\ref{Hamilton})
and, therefore, Hamiltonian $\hat{H}$ are self-adjoined relative to the scalar
product (\ref{scalar-prod}). Hence, the orthogonality of eigenfunctions for
different eigenvalues is provided by the self-adjointness of $\hat{H}$.
However, the quantum number $k$ for current-carrying eigenstates is twofold
degenerate. Thus any complex solution $\psi_{k}$ and its conjugate $\psi
_{k}^{\ast}$ form the linear-independent pair of solutions. In order to the
degenerate pair of eigenfunctions do not violate the necessary orthonormality
of the basis in the Hilbert space generated by Hamiltonian $\hat{H}$ they must
be orthogonalized and normalized.

Let us adopt as $\psi_{k}$ the wave function describing tunneling from right
half-space to the left one and having asymptotes of form%
\begin{equation}
\psi_{k}^{\mathrm{R}}=C_{k}^{\mathrm{R}}\left(  \mathsf{e}^{-\mathrm{i}%
kz}+r_{k}^{\mathrm{R}}\mathsf{e}^{\mathrm{i}kz}\right)  ,\,z\rightarrow
\infty;\,\,\,\psi_{k}^{\mathrm{R}}=C_{k}^{\mathrm{R}}t_{k}^{\mathrm{R}%
}\,\mathsf{e}^{-\mathrm{i}qz},\,\,z\rightarrow-\infty.\label{psi-R asympt}%
\end{equation}
The usual continuity condition of the probability flow density
\begin{equation}
j_{k}(z)=\psi_{k}^{\ast}(z)\left(  \mathsf{\hat{v}}+\mathsf{\hat{v}}%
^{+}\right)  \psi_{k}(z)/2=\mathrm{const}(z)\label{continuity-def}%
\end{equation}
gives the relation between amplitudes of the transmission and reflection
coefficients%
\begin{equation}
\frac{\partial E^{\mathrm{L}}(q,\mathbf{k}_{||})}{\partial q}\left|
t_{k}^{\mathrm{R}}\right|  ^{2}=\frac{\partial E^{\mathrm{R}}(k,\mathbf{k}%
_{||})}{\partial k}\left(  1-\left|  r_{k}^{\mathrm{R}}\right|  ^{2}\right)
.\label{t2-r2 relation}%
\end{equation}
Here $E^{\mathrm{L}}(q,\mathbf{k}_{||})$ and $E^{\mathrm{R}}(k,\mathbf{k}%
_{||})$ are, respectively, the left and right energy spectrum, the velocity
operator is defined by $\mathsf{\hat{v}}=\mathrm{i}\left[  \hat{H},\hat
{z}\right]  $. The conservation of the total energy $E$ and the transverse
momentum $\mathbf{k}_{||}$
\begin{equation}
E^{\mathrm{L}}(q,\mathbf{k}_{||})=E^{\mathrm{R}}(k,\mathbf{k}_{||}%
)=E\label{energy conservation}%
\end{equation}
determines $q(k)$ as a function $k$ and vice versa. Accounting for
Eq.(\ref{energy conservation}), the Eq.(\ref{t2-r2 relation}) can be written
in more compact form%
\begin{equation}
\frac{1}{\partial q/\partial k}\left|  t_{k}^{\mathrm{R}}\right|
^{2}=1-\left|  r_{k}^{\mathrm{R}}\right|  ^{2}.\label{t2-r2-R}%
\end{equation}

The contribution to the normalizing integral (\ref{norm-def}) is formed by
infinite regions of the $z$ axis where the asymptotic expressions (\ref{psi-R
asympt}) are valid. Using the definition (\ref{scalar-prod}) and the relation
(\ref{t2-r2-R}) the value of $\left|  C_{k}^{\mathrm{R}}\right|  ^{2}=1/2\pi$
can be found that provides%
\begin{equation}
\lim_{k\rightarrow k_{1}}\left\langle \psi_{k}^{\mathrm{R}}|\psi_{k_{1}%
}^{\mathrm{R}}\right\rangle =\delta\left(  k-k_{1}\right)  .\label{norm-psiR}%
\end{equation}

The second normalized solution $\psi_{q}^{\mathrm{L}}$ which is
linear-independent of $\psi_{k}^{\mathrm{R}}$ can be obtained by the similar
procedure with the following results%
\begin{equation}
\psi_{q}^{\mathrm{L}}=C_{q}^{\mathrm{L}}\left(  \mathsf{e}^{\mathrm{i}%
qz}+r_{q}^{\mathrm{L}}\mathsf{e}^{-\mathrm{i}qz}\right)  ,\,z\rightarrow
-\infty;\,\,\,\psi_{q}^{\mathrm{L}}=C_{q}^{\mathrm{L}}t_{q}^{\mathrm{L}%
}\mathsf{e}^{\mathrm{i}kz},\,\,z\rightarrow\infty, \label{psi-L asympt}%
\end{equation}%
\begin{equation}
\frac{1}{\partial k/\partial q}\left|  t_{q}^{\mathrm{L}}\right|
^{2}=1-\left|  r_{q}^{\mathrm{L}}\right|  ^{2} \label{t2-r2-L}%
\end{equation}%
\begin{equation}
\left|  C_{q}^{\mathrm{L}}\right|  ^{2}=1/2\pi,\,\,\lim_{q\rightarrow q_{1}%
}\left\langle \psi_{q}^{\mathrm{L}}|\psi_{q_{1}}^{\mathrm{L}}\right\rangle
=\delta\left(  q-q_{1}\right)  ,\,\,q=q(k). \label{norm-psiLq}%
\end{equation}
The other useful normalization
\begin{equation}
\lim_{k_{1}\rightarrow k}\left\langle \psi_{q(k)}^{\mathrm{L}}|\psi_{q(k_{1}%
)}^{\mathrm{L}}\right\rangle =\delta\left(  k-k_{1}\right)  \label{norm-psiLk}%
\end{equation}
is obtained under condition
\begin{equation}
\left|  C_{k}^{\mathrm{L}}\right|  ^{2}=\frac{\partial q/\partial k}{2\pi}.
\label{C4L-k norm}%
\end{equation}

To elucidate a question about mutual orthogonality of $\psi_{k}^{\mathrm{R}}$
and $\psi_{q(k)}^{\mathrm{L}}$ it needs to know the interrelation between the
pairs $\left(  t_{k}^{\mathrm{R}},r_{k}^{\mathrm{R}}\right)  $ and $\left(
t_{q(k)}^{\mathrm{L}},r_{q(k)}^{\mathrm{L}}\right)  .$ The required relations
can be found in general form independently of a particular barrier if one
realizes $\psi_{q(k)}^{\mathrm{L}}$ as the linear combination of $\psi
_{k}^{\mathrm{R}}$ and $\psi_{k}^{\mathrm{R}\ast}$ that gives%
\begin{equation}
t_{q(k)}^{\mathrm{L}}=\frac{1}{\partial q/\partial k}t_{k}^{\mathrm{R}%
},\,\,\,r_{q(k)}^{\mathrm{L}}=-r_{k}^{\mathrm{R\ast}}\left(  t_{k}^{\mathrm{R}%
}/t_{k}^{\mathrm{R\ast}}\right)  .\label{(t,r)L-(t,r)R}%
\end{equation}
Using Eq.(\ref{(t,r)L-(t,r)R}) one can show that terms like $\delta$-function
in scalar product \ $\left\langle \psi_{q(k)}^{\mathrm{L}}|\psi_{k_{1}%
}^{\mathrm{R}}\right\rangle $ cancel each other and hence
\begin{equation}
\lim_{k_{1}\rightarrow k}\left\langle \psi_{q(k)}^{\mathrm{L}}|\psi_{k_{1}%
}^{\mathrm{R}}\right\rangle =0.\label{L-R ortho}%
\end{equation}
The proof of Eq.(\ref{L-R ortho}) completes the construction of the
orthonormal basis in Hilbert space of single-particle Hamiltonian
$\hat{H}$.

\subsection{Electron density and current density}
In the case of an equilibrium system the single-particle density
matrix is $\hat{\rho}=\hat{\rho}(\hat{H})$ and it has only diagonal
non-zero elements $\rho^{\mathrm{LL}}\equiv F^{\mathrm{L}}$ and
$\rho^{\mathrm{RR}}\equiv F^{\mathrm{R}}$ in the chosen basis due to
the orthogonality relation
(\ref{L-R ortho}). Therefore,%
\begin{equation}
n(z)=\mathrm{Sp}\hat{\rho}\hat{n}(z)=\nonumber
\end{equation}%
\begin{equation}
=2\int_{0}^{\infty}dq\int_{-\infty}^{\infty}d\mathbf{k}_{||}F^{\mathrm{L}%
}\left[  E^{\mathrm{L}}(q,\mathbf{k}_{||})\right]  \left|  \psi_{q}%
^{\mathrm{L}}(z)\right|  ^{2}+2\int_{0}^{\infty}dk\int_{-\infty}^{\infty
}d\mathbf{k}_{||}F^{\mathrm{R}}\left[  E^{\mathrm{R}}(k,\mathbf{k}%
_{||})\right]  \left|  \psi_{k}^{\mathrm{R}}(z)\right|  ^{2}.\label{n(z)-exp}%
\end{equation}
Since the Schottky barrier is completely situated in the semiconductor it is
convenient to change the $q$-integration in Eq.(\ref{n(z)-exp}) by
$k$-integration because $k$ is the quantum number of the \textrm{R}%
-eigenstates. The result is%
\begin{equation}
n(z)=2\int_{0}^{\infty}dk\int_{-\infty}^{\infty}d\mathbf{k}_{||}\left\{
F^{\mathrm{L}}\left[  E^{\mathrm{R}}(k,\mathbf{k}_{||})\right]  \left|
\psi_{k}^{\mathrm{L}}(z)\right|  ^{2}+F^{\mathrm{R}}\left[  E^{\mathrm{R}%
}(k,\mathbf{k}_{||})\right]  \left|  \psi_{k}^{\mathrm{R}}(z)\right|
^{2}\right\}  .\label{n(z)-fin}%
\end{equation}
Here the subindex $q(k)$ in $\psi^{\mathrm{L}}$ was changed by $k$ to recall
the necessity to use the normalization (\ref{norm-psiLk})-(\ref{C4L-k norm}).

The diagonal elements $F^{\mathrm{L}}$ and $F^{\mathrm{R}}$ of the density
matrix determine respectively the occupation of \textrm{L}- and \textrm{R}%
-eigenstates and are described by Fermi distributions. The choice of the
\textrm{L}- \textrm{R}-states as the basis allows to account mathematically
for the presence of two independent reservoirs of particles (thermostats) in
the left and right infinities. The Fermi level $E_{F}^{\mathrm{L,R}}$\ of each
reservoir is determined by the proper neutrality condition far away from the
interface and they are different as the bias voltage is applied to the
junction. It is important to stress that both $\psi_{k}^{\mathrm{L}}(z)$ and
$\psi_{k}^{\mathrm{R}}(z)$ states extend to the both infinities and contribute
to $n(z)$ at any point. Thus the bias applied changes the position of each
Fermi level. Substituting asymptotic expressions (\ref{psi-R asympt}) and
(\ref{psi-L asympt}) for $\psi_{k}^{\mathrm{R}}$ and $\psi_{k}^{\mathrm{L}}$
in Eq.(\ref{n(z)-fin}) and using the neutrality condition $n(\infty
)=N_{\mathrm{+}}^{\mathrm{R}}$ we obtain the equation determining the
dependence of the Fermi level of electrons in semiconductor on the bias $U=-V$
(compare with Eq.(3.101) in \cite{Ferry-Good97})%
\[
n(z\rightarrow\infty)\equiv N_{+}^{\mathrm{R}}=\frac{4}{(2\pi)^{3}}\int
_{0}^{\infty}dk\int_{-\infty}^{\infty}d\mathbf{k}_{||}F^{\mathrm{R}}\left[
E^{\mathrm{R}}(k,\mathbf{k}_{||})\right]  +
\]%
\begin{equation}
+\frac{2}{(2\pi)^{3}}\int_{k_{U}}^{\infty}dk\int_{-\infty}^{\infty}%
d\mathbf{k}_{||}\left(  1-\left|  r_{k}^{\mathrm{R}}\right|  ^{2}\right)
\left\{  F^{\mathrm{L}}\left[  E^{\mathrm{R}}(k,\mathbf{k}_{||})\right]
-F^{\mathrm{R}}[E^{\mathrm{R}}(k,\mathbf{k}_{||})]\right\}  .\label{Efermi}%
\end{equation}
Here $V$ is the structure voltage drop, $k_{U}=\left[  2\max(E^{L}%
(0,0)-U,0)\right]  ^{1/2}$, $E_{F}^{\mathrm{L}}=E_{F}^{\mathrm{R}}-U$. At zero
temperature $E_{F}^{\mathrm{R}}=E^{\mathrm{R}}(k_{F})$ and this equation can
be transformed into the equation for the Fermi wave vector of the right
electrons%
\begin{equation}
\frac{k_{F}^{3}}{3\pi^{2}}=N_{\mathrm{+}}+2\mathrm{sgn}(U)\int
\limits_{E(k,\mathbf{k}_{||})\in\lbrack E_{F}-U,E_{F}]}\frac{dkd\mathbf{k}%
_{||}}{(2\pi)^{3}}\left(  1-\left|  r_{k}\right|  ^{2}\right)  .\label{kF}%
\end{equation}
The index \textrm{R} is suppressed here for short. It is easily to see from
Eq.(\ref{kF}) that at $U>0$ we have $k_{F}(U)>k_{F}(0)$ and vice versa as it
should be. Under the low barrier transparency (the reflection coefficient is
of order of 1) the Eq.(\ref{kF}) can be solved by simple iterative
method%
\footnote{In the case of metal-semiconductor junctions the relative
correction to the solution of Eq.(\ref{kF}) due to a shift of the
metal $E_{F}^{\mathrm{L}}$ from the value $E_{F}^{\mathrm{R}}-U$ by
neutrality constraint is of order of $\approx
(U/E_{F}^{\mathrm{L}})(1-|r_{F}^{\mathrm{R}}|^{2})^{2}$ and can be
neglected in calculations of $I-V$ characteristics of real
structures with typical values $1-|r_{F}^{\mathrm{R}}|^{2}<10^{-4}$,
$U<1\,\mathrm{eV}$, $E_{F}^{\mathrm{L}}\lesssim 10\,\mathrm{eV}$.
Direct numeric calculations in Ref.\cite{Mera-eaPRB05} have shown
that a violation of the equality $\Delta E_{F}=U$ becomes essential
when the barrier height is $\lesssim E_{F}$ and barrier width is
$\lesssim 2\pi /k_{F}$.}.

After the Fermi level is found one can calculate the tunnel current, averaging
the current density (\ref{continuity-def}) with density matrix and using the
asymptotic representation of the wave functions at $z\rightarrow\infty$. The
result is%
\begin{equation}
I(U)=-2e\int_{0}^{\infty}\frac{dk}{2\pi}\int\frac{d\mathbf{k}_{||}}{(2\pi
)^{2}}\frac{\partial E}{\partial k}\left[  F(E)-F(E+U)\right]  \left(
1-\left|  r_{k}\right|  ^{2}\right)  . \label{current}%
\end{equation}
Here $F(E)$ is the Fermi distribution at $T\neq0$, $e=-1$ is the electron
charge and $E(k,\mathbf{k}_{||})$ is the energy dispersion relation of the
semiconductor electrons.

\section{Many-electron structures with finite number of electrons}
\label{finite number sec}
\subsection{Many-electron atoms and molecules}
The neutrality problem of iteration procedure in self-consistent
field theory has been considered here for the extended many-electron
systems with undetermined number of electrons and the solution is
suggested in Section \ref{Poiss-iter sec}. In the case of the finite
many-electron systems like atom or molecule the number of electrons
is exactly known and the insolubility of iteration equations seems
not present. Anyway, the required charge state can be prepared if
there is sufficient number of bounded-state levels. However, the
question is then transformed into the impossibility to find the
Poisson equation solution with a given angle symmetry if the charge
density at the right-hand side does not have the desired symmetry.
Fertig and Kohn \cite{Fert-Kohn00}\ has considered the problem and
discussed its source and consequences. But the solution suggested
consists in additional artificial constraints on the sought
variational solution.

It seems, however, that separation of the induced charge in the
self-consistent Poisson equation like it is done in Eq.(\ref{Poiss+Ni}) might
also solve the problem in the case of finite Fermi systems. The induced charge
can be taken in the form like in the Thomas-Fermi-Dirac theory \cite{Bethe64}
with correlation potential included.

\subsection{Periodic electronic structures and supercells}
The spatial periodicity of crystal solids makes possible to replace
numeric simulations of infinitely extended system by computations
for a finite fragment (cell) supplemented by periodic boundary
conditions. To map the preceding analysis of the interrelation
between the charge distribution and boundary conditions for Coulomb
potential on such a case let us rewrite the Eqs. (\ref{1D u'
solution}),(\ref{1D u solution}) for the finite interval
$(a,b)$%
\begin{equation}
u^{\prime}(b)-u^{\prime}(a)=4\pi\int_{a}^{b}dz_{1}\rho(z_{1}%
),\label{1D-u'(a,b)}%
\end{equation}%
\begin{equation}
u(b)-u(a)=(b-a)u^{\prime}(b)-4\pi\int_{a}^{b}dz_{1}z_{1}\rho(z_{1}%
).\label{1D u (a,b)}%
\end{equation}
The periodicity demands the conditions for the total charge $Q\equiv\int
_{a}^{b}dz_{1}\rho(z_{1})=0$ and the total dipole moment
\[
D\equiv\int_{a}^{b}dz_{1}z_{1}\rho(z_{1})=(b-a)u^{\prime}(b)/4\pi
\]
to be fulfilled. In the case of centrosymmetric structure there should be
$u^{\prime}(a)=u^{\prime}(b)=0$ and therefore $D=0$ (see e.g.
\cite{L-L82Electrodyn} \S \S 6, 13). At each $i$th iteration step it is easy
to provide the neutrality condition $Q^{(i)}=0$ for the finite structure with
a prescribed ionic charge by filling the necessary number of electronic states
after solving the Schr\"{o}dinger equation. However, the charge distribution
with $D^{(i)}\neq0$ is the rather likely result that cannot be prevented. In
this case the periodic boundary condition for the solution of the Poisson
equation can be fulfilled only at $u^{\prime}\neq0$ at the boundaries. At the
same time the behavior of the potential should be essential contorted in
comparison with a ''true'' charge distribution at $D=0$ that produces
increasingly distorted charge distribution at the next iteration cycle. In the
case of Fourier-series decomposition of the electron density and potential,
the discontinuity of the potential at the boundaries entails a growth of the
solution $u$ and, therefore, its derivative $u^{\prime}$ again near the
boundaries owing to the Gibbs effect \cite{Zigm59}.

In the case of two- or three-dimensional cell the Eqs. (\ref{1D-u'(a,b)}) and
(\ref{1D u (a,b)}) should be replaced by%
\begin{equation}
\oint\limits_{S_{\mathrm{cell}}}dS\mathbf{\nu}_{s}\mathbf{\cdot}\nabla
u=4\pi\int\limits_{\Omega_{\mathrm{cell}}}d\mathbf{r}\rho(\mathbf{r}%
)\equiv4\pi Q_{\mathrm{cell}},\,\,\,the\,\,Gauss\,\,theorem,\label{Gauss theo}%
\end{equation}%
\begin{equation}
\oint\limits_{S_{\mathrm{cell}}}dS\left[  \mathbf{\nu}_{s}\mathbf{\cdot}\nabla
u-u(\mathbf{\nu}_{s}\mathbf{\cdot}\nabla)\right]  \mathbf{r}=4\pi
\int\limits_{\Omega_{\mathrm{cell}}}d\mathbf{rr}\rho(\mathbf{r})\equiv
4\pi\mathbf{D}_{\mathrm{cell}},\label{D-theo}%
\end{equation}
with the same conclusions as in the one-dimensional case above. Here
$\Omega_{\mathrm{cell}}$ and $S_{\mathrm{cell}}$ are the volume and bounding
surface of the cell, respectively, $\mathbf{\nu}_{s}$ is the external unit
normal to the cell surface. The formulae (\ref{Gauss theo}) and (\ref{D-theo})
are derived from the Poisson equation with the use of the Green theorem%
\begin{equation}
\Delta u(\mathbf{r})=4\pi\rho(\mathbf{r}),\,\,\,\,\int_{\Omega}d\mathbf{r}%
\Phi\Delta\Psi=\int_{\Omega}d\mathbf{r}\Psi\Delta\Phi+\oint_{S_{\Omega}%
}d\mathbf{S}\left(  \Phi\nabla\Psi-\Psi\nabla\Phi\right)  \label{Green theo}%
\end{equation}
that requires for proper differentiability of the involved functions and
correspondent smoothness of the bounding surface $S$ to be fulfilled
\cite{Vlad71-EqMatPhys}.

The incompatibility of the boundary conditions with the right-hand-side of the
Poisson equation makes the Dirichlet problem as ill-posed one with
corresponding strong perturbance of the solution in response to insignificant
perturbance of the initial data \cite{Courant62}. This mechanism may be
responsible for the observed deterioration of current self-consistent
iterative algorithms manifested in the long-wavelength charge instability,
which is named as ''charge sloshing'' \cite{KerkerPRB81}%
-\cite{Kress-FurthPRB96}.

It is necessary to note that at initial iteration step the solution of
self-consistent Poisson equation (\ref{Poiss+Ni})\ with the induced electron
distribution $n_{\mathrm{ind}}$\ defined by Eq.(\ref{Ni-quasicl}) should give
a good starting guess for the potential and the valence electron distribution
in the cell since they are self-consistent and meet the boundary conditions.
However, the expression (\ref{Ni-quasicl}) is evidently inappropriate in the
case of crystals with the filled energy bands and it needs to use a next
quasi-classical approximation for the induced electron density expressed as a
function of derivatives of the potential (see, e.g. \cite{Kirzh63}).

Let us consider now the application of the results of Sec. \ref{Schred-Hilb
Sec} to the continuum spectrum of infinitely extended periodic system. The
eigenfunctions of single-particle Hamiltonian according to Bloch theorem can
be taken in the form%
\begin{equation}
\psi_{\mathsf{j}\mathbf{k}}(\mathbf{r})=C\exp(\mathsf{i}\mathbf{kr}%
)\phi_{\mathsf{j}\mathbf{k}}(\mathbf{r}), \label{Bloch w-f}%
\end{equation}
where $\phi_{\mathsf{j}\mathbf{k}}(\mathbf{r})$ is the cell-periodic part,
$\mathsf{j}$ is the energy band index, $\mathbf{k}$\ is the wave vector, and
$C$ is the normalizing constant. The definition of the scalar product like in
Eq.(\ref{scalar-prod}) by the expression%
\begin{equation}
\left\langle \psi_{\mathsf{j}\mathbf{k}}|\psi_{\mathsf{j}_{1}\mathbf{k}_{1}%
}\right\rangle =\lim_{\epsilon\rightarrow0}\int d\mathbf{r}\exp(-\epsilon
|\mathbf{r}|)\psi_{\mathsf{j}\mathbf{k}}^{\ast}(\mathbf{r})\psi_{\mathsf{j}%
_{1}\mathbf{k}_{1}}(\mathbf{r}) \label{norm-def Bloch}%
\end{equation}
and the eigenfunction norm like in Eq.(\ref{norm-def}) provides the
self-adjointness of single-particle Hamiltonian and results in the orthonormal
basis of the Hilbert space with $|C|^{2}=(2\pi)^{-3}$%
\begin{equation}
\left\langle \psi_{\mathsf{j}\mathbf{k}}|\psi_{\mathsf{j}_{1}\mathbf{k}_{1}%
}\right\rangle =\delta\left(  \mathbf{k}-\mathbf{k}_{1}\right)  \delta
_{\mathsf{jj}_{1}} \label{ortho-norm Bloch}%
\end{equation}
under natural conditions $\mathbf{k},\mathbf{k}_{1}\in$ 1st Brillouin zone
and
\begin{equation}
\int\limits_{\Omega_{\mathrm{cell}}}d\mathbf{r}\left|  \phi_{\mathsf{j}%
\mathbf{k}}(\mathbf{r})\right|  ^{2}=\Omega_{\mathrm{cell}}. \label{norm Fi-Bloch}%
\end{equation}
The relationships just obtained allow to determine the electron
density inside the cell by
\begin{equation}
n(\mathbf{r})=2\sum_{\mathsf{j}}\int\limits_{\mho_{\mathrm{BZ}}}%
\frac{d\mathbf{k}}{(2\pi)^{3}}\left|  \phi_{\mathsf{j}\mathbf{k}}%
(\mathbf{r})\right|  ^{2} \Theta(E_{F}-E_{\mathsf{j}}(\mathbf{k})), \label{n(r)-cell}%
\end{equation}
where the multiplier $2$ takes into account the spin degeneration and
$k$-integration is taken over the unit cell of the reciprocal lattice that is
the first Brillouin zone. Because the volume of the Brillouin zone
$\mho_{_{\mathrm{BZ}}}=(2\pi)^{3}/\Omega_{\mathrm{cell}}$ we have%
\begin{equation}
\int\limits_{\mho_{\mathrm{BZ}}}\frac{d\mathbf{k}}{(2\pi)^{3}}=\frac{1}%
{\Omega_{\mathrm{cell}}}. \label{dk-Brill-volume}%
\end{equation}
Thus the band filling and the position of the Fermi level can be determined
from the neutrality condition%
\begin{equation}
\int\limits_{\Omega_{\mathrm{cell}}}d\mathbf{r}n(\mathbf{r})=N_{\mathrm{+}},
\label{dr-n(r)-cell}%
\end{equation}
where $N_{\mathrm{+}}$ is the ion charge of the unit cell.

The expressions (\ref{norm Fi-Bloch})-(\ref{dr-n(r)-cell}) allow to abandon
the use of unnecessary Born-Karman boundary condition for the eigenfunctions
$\psi_{\mathsf{j}\mathbf{k}}(\mathbf{r})$, which limits admissible points in
$k$-space by the discrete set. As a result, one can calculate the energy bands
$\varepsilon_{\mathsf{j}}(\mathbf{k})$ of perfect crystal or make integrations
over Brillouin zone, using any set of $k$-points dictated by selected
algorithm \cite{Monk-PackPRB76}, and avoid disadvantages of the cell size
extension beyond the size of the minimal unit cell to increase the sampling
density of $k$-points \cite{Niem-PRB05}.

Similarly, the artificial periodicity of the supercell \cite{Payne-TeterRMP92}%
\ can be eliminated from the calculations of solid surface or imperfect
crystal with a point defect and replaced by asymptotic boundary conditions at
infinity for the Coulomb potential and wave functions. This removes the known
difficulties introduced in calculations of extended systems by making use the
slab \cite{App-HamRMP76} or supercell \cite{Mak-PayPRB95}%
-\cite{Niem-PRB05-Supcell}\ geometries. In this case the induced charge
resulted from the free carriers or nonuniform polarization of the valence
electrons should be introduced into the self-consistent Poisson equation using
the effective mass approximation or the macroscopic electric susceptibility, respectively.

\section{Concluding remarks}
The origin of all difficulties with self-consistency listed above is
the long-range character of the Coulomb interaction that is
responsible for an interdependence of the charge distribution with
the boundary conditions on the potential. 
Thus large-scale self-consistent distributions can only result from
the direct solution of the self-consistent Poisson equation itself
because the boundary conditions take into account the reaction of
distant charges that are exterior ones relative to the considered
system. The separation of the induced charge as a function of the
potential and the modification of the Poisson equation is only way
in order to obtain the self-consistent distributions of potential
and 
charge with due accounting for the boundary conditions.

To elucidate the point in more detail let us consider the electrostatic part
of the Kohn-Sham energy functional%
\begin{equation}
E_{\mathrm{es}}=\frac{1}{2}\int d\mathbf{r}d\mathbf{r}^{\prime}\frac
{\rho(\mathbf{r})\rho(\mathbf{r}^{\prime})}{\left|  \mathbf{r-r}^{\prime
}\right|  } \label{E_es}%
\end{equation}
and the corresponding contribution to the effective potential in the
Schr\"{o}dinger equation \cite{K-V83}%
\begin{equation}
u_{\mathrm{es}}=\int d\mathbf{r}^{\prime}\frac{\rho(\mathbf{r}^{\prime}%
)}{\left|  \mathbf{r-r}^{\prime}\right|  }. \label{u_es}%
\end{equation}
At first glance, there is no need for the Poisson equation since Eqs.
(\ref{E_es}) and (\ref{u_es}) give us already the explicit relationships
between the necessary quantities and the charge density. However, the function
$1/\left|  \mathbf{r-r}^{\prime}\right|  $ in the integrand of the Eq.
(\ref{u_es}) is the Green function of the Laplace equation with the zero
boundary condition at the infinity. The expression (\ref{u_es}) is the true
solution of the Poisson equation if there are no charges outside the
integration region. Evidently, this is not the case for infinitely extended
systems or when periodic boundary conditions are specified.

Let us assume that there are two subsystems with non-intersected charge
distributions 
\begin{equation}
\rho(\mathbf{r})=\rho_{1}(\mathbf{r})+\rho_{2}(\mathbf{r});\,\rho
_{1}(\mathbf{r})\neq0,\,\mathbf{r\in\Upsilon}_{1};\,\rho_{2}(\mathbf{r}%
)\neq0,\,\mathbf{r\in\Upsilon}_{2};\,\mathbf{\Upsilon}_{1}\cap\mathbf{\Upsilon
}_{2}=\emptyset\label{ro-2}%
\end{equation}
and, respectively,%
\begin{equation}
\varphi(\mathbf{r})=\varphi_{1}(\mathbf{r})+\varphi_{2}(\mathbf{r}%
);\,\Delta\varphi_{1,2}=-4\pi\rho_{1,2}, \label{fi-2}%
\end{equation}
where $\varphi$ is the Coulomb potential. Substituting the expression
(\ref{ro-2}) for $\rho$ in Eq. (\ref{E_es}) and using Poisson equations
(\ref{fi-2}) together with Green theorem (\ref{Green theo}) we obtain%
\begin{equation}
E_{\mathrm{es}}=\int\limits_{\mathbf{\Upsilon}_{1}}d\mathbf{r}\rho_{1}%
\varphi+\frac{1}{8\pi}\int\limits_{\mathbf{\Upsilon}_{1}}d\mathbf{r}%
\varphi\Delta\varphi+\int\limits_{\mathbf{\Upsilon}2}d\mathbf{r}\rho
_{2}\varphi+\frac{1}{8\pi}\int\limits_{\mathbf{\Upsilon}2}d\mathbf{r}%
\varphi\Delta\varphi\equiv E_{\mathrm{es}1}+E_{\mathrm{es}2}. \label{E_es1+2}%
\end{equation}
Now it is possible to consider the total energy $E_{\mathrm{tot}1}$
of the
subsystem 1 as the functional of $\rho_{1}(\mathbf{r})$, $\Psi_{E1}%
(\mathbf{r})$, and $\varphi(\mathbf{r})$. Then the necessary conditions of the
functional minimum are \cite{K-V83}%
\begin{equation}
\delta E_{\mathrm{tot}1}/\delta\Psi_{E1}^{\ast}(\mathbf{r})=0,\,\delta
E_{\mathrm{es}1}/\delta\varphi(\mathbf{r})=0\label{variations}%
\end{equation}
that must be supplemented by the expression for the effective potential
$u_{\mathrm{eff}1}\equiv\delta E_{\mathrm{es}1}/\delta\rho_{1}+u_{\mathrm{xc}%
1}$. The first equality gives rise to the Schr\"{o}dinger equation and the
second one is just the Poisson equation. The variation $\delta\varphi$ and the
sought potential $\varphi$\ have to satisfy such boundary conditions that
ensure%
\begin{equation}
\oint d\mathbf{S}\left(  \varphi\nabla\delta\varphi-\delta\varphi\nabla
\varphi\right)  =0\label{dS=0-condition}%
\end{equation}
at the surface confining the region $\mathbf{\Upsilon}_{1}$ only. It
implies that boundary conditions to Poisson equation must be
properly fixed during the iterative process. If a realization of the
equality (\ref{dS=0-condition}) at given boundary conditions turns
out to be impossible then the separate investigation of two
subsystems is incorrect. This derivation shows that the solution of
the Poisson equation cannot be replaced by direct variations of Eq.
(\ref{E_es})-(\ref{u_es}) in the course of searching for the
$E_{\mathrm{tot}1}$\ minimum for the infinitely extended electronic
systems with the periodic or assigned asymptotically at infinity
boundary conditions.

It is desirable to add some comment on the widely-used ''mixing'' method of
fight against the charge instability. The very essence of the method lies in
the use of some linear combination of results of previous iterative steps to
make up the input for the next step. Such an approach is named as ''iteration
with memory'' in the iterative calculus and it is destined to accelerate the
convergence rate of the iterative process \cite{Traub82-Iterative}. But this
procedure cannot transform a divergent iteration scheme to convergent one.
There are many reasons, including mentioned in the present work, to believe
that charge instability observed in simulations of large electronic structures
is rather sign of divergency than slow convergency of the simple iteration
cycle of Poisson$\rightarrow$Schr\"{o}dinger$\rightarrow$Poisson steps. The
continuous appearance of new more and more cumbersome and sophisticated mixing
methods during the last three decades is the best evidence of this point of
view (see some historic commentary in \cite{Liebsch97}, \cite{KerkerPRB81}%
-\cite{Kress-FurthPRB96}, \cite{Goed99scaling}). As a rule, the
demonstration of the benefit of a new method is accompanied by
illustrations of the lack of convergence of the old one as the
structure size becomes larger. It is of interest, the mixing scheme
based on some handmade modeled screening succeeds relatively
\cite{Goed99scaling}, although it was constructed on the ground of
rather formal mathematical reasoning than the underlying physics
discussed here.

It needs also to note that mixing schemes may produce a spurious convergency
(see \cite{Kress-FurthPRB96}, p. 11176). Thus it is necessary to check whether
the norm of the functional derivatives $\left\|  \delta E_{\mathrm{es}%
1}/\delta\varphi(\mathbf{r})\right\|  $ and $\left\|  \delta E_{\mathrm{tot}%
1}/\delta\Psi_{E1}^{\ast}(\mathbf{r})\right\|  $, which are residuals of
Poisson and Schr\"{o}dinger equations, are minimal along with\ the total
energy $E_{\mathrm{tot}1}$. Of course, $E_{\mathrm{tot}1}$ is the total energy
of sufficiently large but finite system that can be approximated by assignment
of boundary conditions as in the infinitely extended one.

\ack
I am indebted to Prof. A. Liebsch who has paid my attention to a
neutrality problem in iterative solution of the self-consistent
field equations and to Dr. H. Mera for useful discussion of
computational details pertinent to Ref.\cite{Mera-eaPRB05}. The
partial financial support by Russian Foundation for Basic Researches
and die Deutsche Forschungsgemeinschaft is acknowledged.

\end{document}